\documentstyle[12pt,psfig]{article}
                                                       
\begin{document}                                                                
                                                                                
\def \vsp {\vspace{.4cm}}
\def \vss {\vspace{.4cm}}                                                       
\def \vsl {\vspace{1cm}}                                                        
\def \fl {\flushleft}                                                           
\def \fr {\flushright}                                                          
\def \S  {$\sqrt s$ }
\def \xf {$x_F$ }
\def \dsdn {$d\sqrt s/ dn$ }
\def \pt {$p_t$ }

\def \LAB {$A^{1/3} + B^{1/3}$ }
\def \LAC {$(N-1)/2 + (M-1)/2 $ }
\def \jps {$J/\Psi$'s }
\def \soc {$\sigma_{oc}$ }
\def \jp {$J/\Psi$ }
\def \Et {$E_t$ }
\begin{center}
{\bf \jp Suppression Revisited

\vss
S. Frankel and W. Frati

\vsp
Physics Department, University of Pennsylvania

\vsl

    Abstract }

\end{center}

      The production of \jp's in nuclei is re-examined
 and the critical role of Feynman x distributions in the study of
photoproduction and hadronic production
of both \jps and Drell-Yan dimuons
           is expatiated. The need to consider both initial and final
state interactions and the key effect of initial state interactions
on the total \jp cross section are  demonstrated.

We first present a theoretical study of the
expected {\it functional form} for the projectile and target
atomic number (A,B)
 dependence of the \jp suppression, S,
 and
demonstrate why it cannot fall exactly exponentially with
 $A^{1/3} + B^{1/3}$, since such a term is multiplied by a weak
{\it enhancement} factor.
  We
                  use  a Woods-Saxon
Monte Carlo simulation to obtain distributions of the numbers of
collisions prior and subsequent to the charmonium production. 
With those results we relate the mean number of nucleon-nucleon
collisions, $<n>$ to
\LAB, and give a new analytic functional form
 for the A dependence of the
suppression.

Finally we  carry out a full Monte Carlo calculation of S, including an
initial state prior energy loss
in keeping with  the measured Feynman \xf
 distributions in $\pi$-A and p-A
interactions. We use an energy loss parameterization
which is  consistent with the energy loss from minimum bias data fitted
in ISAJET. We also use an open charm absorption cross section, \soc,
                                 constrained from photoproduction of
\jps, to analyze the latest data on the \jp suppression.

 The
role of color screening, the time evolution of off-shell charmonium
 to the on-shell \jp, and
the energy dependence of the charmonium-nucleon inelastic and absorption
cross sections are discussed. A program of related measurements, necessary
for better understanding the \jp suppression, is presented.

\vsp  We find
 no anomalous
suppression when comparing our results to    published data.

\newpage
\centerline{\bf I. Review}

\vsp Shortly after the  proposal \cite{Matsui}
of the possibility
of seeing evidence for deconfinement of charmonium in nuclear reactions,
use was made of studies of soft processes in relativistic nuclear
reactions \cite{Predicting}  
 to examine the rich Feynman x distributions of \jp's
that had been measured in production of \jps in pion and proton
interactions in nuclei.
  From the
 outset \cite{jpsi},
charmonium absorption, as the
{\it only} source of \jp suppression, was ruled out by a glance at the
 many
examples of the strong \xf dependence of the p-A/p-p and $\pi$-A/$\pi$-p
ratios of the
\jp yields. In that first work, only               the energy loss of
incoming hadrons before the charmonium production and energy loss due to
rescattering of charmonium
off nucleons in the final state were considered. A more careful
 examination
of many examples of \xf distributions for both \jp and Drell-Yan 
production was then carried out \cite{dimuon}.

 A direct
determination of the charmonium absorption cross section, $\sigma_{oc}$,
from photoproduction of \jp 's was then
obtained \cite{absorption}. In photoproduction
there are no  initial state interaction effects to confuse the
interpretation, as in production from hadron probes.
       We obtained the
result,  \soc = 6.6 +/- 2.2 mb.

\vsp
It is important to understand that there is, presently,
no way to calculate the
non-perturbative effect on the nucleon structure functions resulting
from prior soft scatters before the high $Q^2$ production of 
vector mesons.
Thus the ``energy loss'' must be treated as a parameter that can be
obtained, in principle, from analysis of the Feynman x distributions.

\vsp
 To carry out
 these early calculations we used the well-known
ISAJET model of Frank Paige to arrive at an expression for the
approximate functional form of the  \S dependence of the
energy loss in soft collisions prior to the \jp production.
 The energy loss per collision
is found in low \pt interactions to vary approximately as \dsdn = a +b\S.
The functional form is useful but the values of $a$ and $b$ obtained from
ISAJET are only a guide. For many calculations of the \xf distributions
it is sufficient to use \dsdn = a alone.
      We
   studied \dsdn  in the range .4 to 1.0 in our early work. (Ref.
\cite{dimuon}

 In  Ref. \cite{absorption},
 we presented two key examples of our earlier
 calculations of
the Feynman x distributions for a pion- and  a proton-induced reaction.
Two samples of data
  are reproduced in Fig. 1, since they clearly   illustrate the many
effects that are responsible for \jp suppression:

 \vsp
 \begin{figure}[h]
 \centerline{\psfig{figure=jpffig1,height=12cm}}
 \caption{}
 \end{figure}
 \vsp


\vsp
 1) Loss of \jp's due to
absorption: It is generally assumed,
            since there are no data to test the hypothesis, 
 that \soc is energy
independent. Thus the charmonium absorption  alone
can only reduce $S(x_f)$
{\it independent} of \xf. A flat reduction independent of \xf is not
what the data show. All measured reactions show falloff of the ratio
of the nuclear to nucleon production with \xf.
Further, suppressions at \xf = 0 are not the same for
 pion and proton induced reactions, additional proof that final
 state
absorption cannot be the only mechanism for the suppression.

\vsp
 2) Initial state energy loss: This has two effects: a)
the yield is reduced because of the exponential dependence of the \jp
yields
on 1/\S, $e^{-M/\sqrt s}$,
and b) events are shifted to lower \xf. Both effects
appear in all the measured \xf distributions.

\vsp
 3) The outgoing charmonium interacts
with nucleons inelastically,
losing further energy by soft collisions: \jp + n
 $\rightarrow$ \jp + X .
This will affect
the shape of the \xf distributions as well, forcing events to lower \xf.

\vsp
(In Fig. 1 the absorption cross section was taken as 9 mb and \dsdn taken
as .4 but a smaller cross section and somewhat larger \dsdn would provide
a similar fit. Also,
          an energy loss
of the outgoing \jp of $\Delta E/E$ = .1 resulted in a need for an 
inelastic cross section, $\sigma_{inel}$, of about 11 mb.)

\vsp
Clearly the Feynman x distributions are complicated, but they are rich in
information.    Fortunately the inelastic \jp interactions will not affect
the total \jp suppression ratios.
 (However, they will affect the rapidity distributions and thus
   affect the rapidity acceptance, producing a possible
experimental
 A dependent effect on S.)

\newpage
\centerline {\bf II. Bases for Suppression Calculations }

    \vsp

  We digress to emphasize some of the constraints that time dilation
places on multiple scatterings in nuclei: Immediately after a low \pt
proton collision on a nucleon in a nucleus, the proton is in an
``excited'' state. It is
`` off-shell''. The
asymptotic final state of the  proton is a real ``on-shell''
proton. Real pions and other produced mesons appear later in the
asymptotic state.
      The time for this evolution in the proton center of mass
 is usually assumed to
be roughly set by the uncertainty relation, $\Delta E \Delta t = \hbar$.
t is then time dilated so that the final ``asymptotic''
state of on-shell proton and pions occurs much later. 

(One would like to believe this argument is
quantitative, but the time evolution is not presently understood, as
recent studies of color screening have indicated \cite{physlett92}.

For light nuclei
it is known that the hadronization
     occurs when the hadrons have left the nucleus \cite{physlett82}.
Since the measured multiplicity distributions in p-D, p-He, and He-He
interactions are well fit from the p-p multiplicity distributions and
the probabilities of making n wounded nucleons, neglecting any final state
effects is reasonable for light nuclei. However there may be final state
corrections needed for the multiplicity or \Et distributions in heavy
nuclei. Thus, one cannot use the p-p measured \Et distributions directly
in predicting the nuclear \Et distributions.

  In this picture it is the excited off-shell proton, $p^*$,
which
collides with nucleons to produce the charmonium
in the following scatterings. We have made, in
all our calculations, the assumption that the $p^*$ - nucleon cross
 section
is little different than the p-nucleon cross section, on the possible
ground that high energy nucleon-nucleon cross-sections are essentially
geometric, as the energy independence of the p-p cross sections suggest.
However, if the cross sections were to increase as the protons become
excited, calculations of the number of soft collisions, prior and 
subsequent to the charmonium production, would be enhanced, increasing
the energy loss effect. It is useful to appreciate this difference
in the low \pt collisions in p-A and A-B collisions.

  However, in A-B interactions, the charmonium
may be interacting with nucleons struck by other incoming nucleons, so
the A-B absorption is not on ground state nucleons but sometimes on
excited nucleons.  This too is ignored in the model but it points out
that \jp production in p-A reactions cannot strictly
be treated identically with
A-B interactions.

      The high $Q^2$ charmonium state, however,
 has a short lifetime, so that $\Delta t$ is much smaller than in low \pt
collisions. Therefore it
    is believed to be mainly
made within the nucleus where it can interact with nucleons
further along the path and be absorbed by reactions leading to open 
charmed particles.

Because the pions and kaons made in the soft interactions mainly
materialize
outside the nuclei, these ``co-movers'' should not appreciably
affect the suppression.
We do not consider them and we shall see that they are not needed to
account for the measured suppressions, although they might have a
small
effect on S.
(Aspects of this subject are treated
theoretically, and data on pion production by muons has been examined
 in a
study of the x dependence of pion production in nuclei.)\cite{dependence}
``A Dependence of Hadron production in Inelastic Muon Scattering and
Dimuon Production by Protons'', S. Frankel and W. Frati, Phys. Rev. D51
4783 (1995).

It is clear that the soft energy loss of the pertinent {\it constituents},
gluons and quarks, which produce \jp's and Drell-Yan pairs, and which
should be different for proton and pion projectiles, will have to be
parameters obtained from the Feynman x distributions.

In our work Ref. \cite{dimuon},
  studying \dsdn for \jp's, we observed empirically that the
value of \dsdn 
 was smaller,
$\cong$ .2, for Drell-Yan pairs. This latter value is very crude but
provided a rough fit to our analyses of various early experiments
\cite{badier}.

                      As shown in Fig 1,
reasonable fits to both 800 GeV proton data and the 150
Gev pion data with a charmonium-nucleon inelastic cross section of
about 11 millibarns were obtained, assuming no color transparency
effects were present.
(It was assumed in these analyses
that no color screening was present and that \soc was
\xf independent.)

 It was abundantly clear that energy loss in
the initial state as
well as the absorption into open charm played important roles.
The initial state energy loss was crucial to the \jp analyses and was
small but not zero in the Drell-Yan data.

\vsp
     Thus, in this paper we
        return to examining the new data on \jp suppression in
heavy nuclei since the recent analyses \cite {kharzeev} 
 make assumptions which
 the
Feynman x data contradict.

\newpage
\centerline{\bf III. Analytic Study of the A Dependence of the
\jp Suppression }

\vsp  In this section we see what insights can be obtained from
 collision of a {\it 2 lines} of interacting nucleons to obtain
simple formulas for the effects of energy loss and open charm
absorption on \jp and dimuon production.
As shown in Fig. 2. $N$ and $M$ are the number of nucleons
in each nucleus along a collision line and $n$ and $m$ are the nucleon
positions in
the line.
In a nucleus-nucleus collision there will be a distribution of such
line-on-line collisions and the final result will require averaging over
the $N + M$ distribution. We shall describe such Monte Carlo results later in
this paper but the analytic calculation for the line-on-line expression
for the \jp suppression contains  interesting information.

\begin{figure}[h]
\centerline{\psfig{figure=jpffig2,height=4cm}}
\caption{}
\end{figure}

  We shall assume for this calculation that all the
measurements are taken at the same projectile energy so that $e^{-\alpha}$
represents the reduction in yield per soft collision due to energy loss
{\it prior} to
the \jp production and $e^{-\beta}$ represents the absorption loss in a
single \jp-nucleon collision {\it subsequent} to the production.
 (No difference in any of the cross sections due to proton-neutron
differences appears in this
simple
analysis, although it will affect yields at high \LAB .)

\vsp Consider the \jp production in the scattering of nucleon
$n = 2$ on nucleon $m = 3$
as shown in Fig. 2: $n = 2$ is slowed down by soft interactions with
$m = 1$ and $m = 2$, while $m = 3$ has been slowed by interaction with $n = 1$.
Our calculation sums over the $n, m$ variables. Similarly, nucleons 3-6 in $N$
and 4-7 in $M$ interact with the produced charmonium, these collisions
absorbing the charmonium into
free charmed particles, with an absorption cross-section denoted as \soc.
Because of the high $Q^2$ (low cross-section)
nature of charmonium production,
the number of \jps produced is proportional to NM, all nucleons being
equally likely to interact to make a \jp.

\vsp The energy loss effect comes from the strong \S dependence of the
\jp cross section which is known to be  given by:

\vsp
1)
$ \sigma_{J/\Psi} = e^{-\gamma M/\sqrt s }$

\vsp
with  $\gamma$ =  14.5 and M = 3.1 GeV.

It is important to note that this expression
is \S dependent and the measurements
cover a wide variety of \S.

\vsp  Taking into account prior energy loss,
this becomes:

\vsp
2) $ e^{-\gamma M/(\sqrt s_0 -A n)} = e^{-\gamma M/(\sqrt s_0 \times
[1 -A n/\sqrt s_0]) }         $ 

\vsp
 where A = \dsdn and $n$ is the number of
collisions prior to the \jp production..

Expanding the denominator, it
      can be approximated by
$ e^{-(\gamma M/\sqrt s)  (1 + A n/ s_o)}$.

Thus we have the simple {\it linear approximation}:

\vsp
2') $e^{-\alpha n} \cong e^{-(\gamma M A/ s_o )n } $.

\vsp
(Note that the energy loss effect depends on s and not \S as in the cross
section for \jp production and that
the lower the energy of the reaction, the
larger the suppression.)

As we shall see, this is a fair approximation and we shall use it in our
analytic studies.
But we shall use the exact formulation
in our Monte Carlo calculations  reported later in this paper.

Thus, for the interacting nucleons shown in Fig. 2, we have:

\vsp
3) energy loss: $e^{-\alpha[(n-1) + (m-1)] } $

\vsp

Similarly we can parameterize the loss into open charm where $\beta$ is
proportional to the open charm cross section, $\sigma_{OC}$, assuming that
$\sigma_{OC}$ is energy independent.

\vsp
4) open charm: $e^{-\beta[(N-n) +(M-m)]}$

\vsp
This  product can be re-factored as

open charm: = $ e^{-\beta [(N-1) + (M-1)] } \times e^{+\beta [(n-1) +
      (m-1)]}$

\vsp
5) Including both energy loss and open charm we obtain:

\vsp
 $  S_p  =  e^{-\beta [(N-1) + (M-1)] } \times
e^{-(\alpha -\beta)  [(n-1) +(m-1)]}\times 1/NM $

\vsp Summing over the nucleons,
this can be rewritten as:

\vsp
6) $  S_p  =  e^{-\beta (N+M)} \times \sum_1^N \sum_1^M
      e^{-(\alpha - \beta)(i-1) } e^{-(\alpha - \beta) (j-1) } $

\vsp  Carrying out the sums and rearranging, we get the final result:

 \vsp

7) $ S_p   = e^{- (\alpha + \beta) (N-1 + M-1)/ 2 }$ $\times$
\begin{large}
       $ [ \frac {sinh (\alpha - \beta)N/2}{  N sinh (\alpha - \beta)/2}]
[ \frac {sinh (\alpha - \beta)M/2}{  M sinh (\alpha - \beta)/2}  ]$
\end{large}

\vsp
(N-1 + M-1)/2  can be considered the {\it mean number }
of prior or subsequent
collisions in the row. We shall see later how this is related to
the mean number of collisions calculated from a Woods-Saxon distribution
which we relate to \LAB .

We denote the bracketed factor
in 7) as $C$. Since it increases with $N$ and $M$,  $C$
is an {\it enhancement
factor} as opposed to the exponential suppression factor it multiplies.

Note that the 
{\it cancellation} of $\alpha$ and $\beta$ in $C$  helps in making
$C$ small and
the exponential fall-off a better approximation.

We will return to this effect later when we discuss central $A-B$
collisions,  where the enhancement effect can be seen to be an appreciable
contribution to the $A$ dependence of $S$.

A useful approximation for C  is:

\vsp
8) $ C \cong
 [ 1 + (\alpha - \beta)^2 (N^2-1)/6 ]
       [ 1 + (\alpha - \beta)^2 (M^2-1)/6 ]  $

\vsp Thus we have demonstrated
     that the cross section does indeed fall off
exponentially with
$\alpha  + \beta$ {\it
except} for the enhancement  factor, C,  given by the sinh
expression. 
 The
enhancement effect is understood by recognizing that the interactions
at the very front of the nuclei have only open charm absorption while
the interactions at the very rear of the nuclei have only energy loss.
However, at other positions the effects are both present.

\vss We now turn to a comparison of \jp production to Drell-Yan
production. In this case there is no open charm contribution for the
Drell-Yan production, only
energy loss entering into the ratio, denoted as $S_{dy}$. The result is:

\vsp 9) $S_{dy}
=   e^{- (\alpha -\alpha^\prime ) (N-1 + M-1)/ 2 }$ $\times ~~ C_{dy}$,
where

$ C_{dy} = $
\begin{large}
   $ [ \frac {sinh (\alpha -\beta )N/2}{  N sinh (\alpha -\beta ) /2}]
      [ \frac {sinh (\alpha -\beta )M/2}{  M sinh (\alpha -\beta ) /2}  ]$
   $  \left        / [ \frac {sinh (\alpha^\prime  )N/2}
{  N sinh (\alpha^\prime /2 )}]
      [ \frac {sinh (\alpha^\prime M/2}
{  M sinh (\alpha^\prime /2 )}  ]   \right.  $

\end{large}

\vsp
In this expression $\alpha^\prime$ is the Drell-Yan energy loss parameter
which  is considerably smaller than that for \jp production. We have
estimated \dsdn = 0.2 in our prior studies of Drell-Yan production in
nuclei. \cite{dimuon} \cite{absorption}

\vsp  {\it
Prediction: The ``slopes'' of the logarithmic plots of $S_p$ vs the mean
number of scatters should be somewhat larger than that for
 $S_{dy}$.}

\vss We now turn to an extension of the analytic calculation, which
depends on the mean number of scatters in a line of interactions, to
integration of the line interactions over the spatial configurations
of the nuclei with the goal of relating \LAC to \LAB.
We have shown that the \jp suppression that would take place if a string
of N nucleons interacted with a string of M nucleons along a line is
given approximately by

\vsp
10) $ S = e^{-(\alpha + \beta)[\frac{(N-1) + (M-1)}{2} }    $

\vsp The $C$ of equations 7, 8
is a function of $\alpha$ , $\beta$ , $N$ and $M$ but is close to
unity so we will ignore it for the present discussion. In this form
$\beta$ represents the loss due to \jp absorption but $\alpha$ represents
an approximation of the effect of prior energy loss of the incoming
nucleons on the \jp yield.

\vsp The question we address here is how $\frac {(N-1) + (M-1)}{2}  $
depends on the quantity $ K (A^{1/3} + B^{1/3} -2) $ since this is the
variable that many workers use in plotting the suppression data.
(We choose to keep the -2 in these expressions since,
for p-p collisions, these quantities are zero,
so $S$ is unity.)

\vsp It is clear that equation 10) would have to be integrated over the
probability that one gets N and M nucleons in a line, averaged over
the impact parameter in a nucleus-nucleus collision.

\vsp 
Another question is whether the needed average is some simple function of
 $  (A^{1/3} + B^{1/3}- 2 ) $.

To answer these questions we do not (at first)
 go to a full scale Monte Carlo calculation of the
\jp suppression but first consider only the distribution of prior and
subsequent collisions, $ n_{prior} = n_{subs} = n $,
 before and after the \jp production.

These are needed to examine eq. 7.
We  obtained these  distributions
from a full Monte Carlo, using Woods-Saxon representation of the nuclei.

The averages
for p-A and A-A interactions are shown in Fig. 3. In this figure $n_A +
n_B$ is the sum of the prior scatters in both nuclei.

\vsp
\begin{figure}[h]
\centerline{\psfig{figure=jpffig3,height=8cm}}
\caption{}
\end{figure}
\vsp
(For deuterium and helium we
have used the best spatial distributions, as was done in our search for
deconfinement in p, d, and $\alpha$ experiments at the ISR
cite{akesson}.
This shows up in the slight curvatures at very low A.)

\vsp
The calculations show that the slope of $<n>$ vs $  (A^{1/3} + B^{1/3}) $
is essentially
the same for p-A and A-A interactions, and a little reflection will
allow the reader to realize that this is what is to be expected.
We can then ask what value of K is needed to calculate $<n>$. What we
find from the slopes in  the figure is that K $\cong  .46$.

\vsp 

The fits to a straight line, $<n>$ = $a$  +b(\LAB -2),
                        are  given in Table 1.

\centerline{ Table 1.}
\begin{table}[h]
\begin{center}
\begin{tabular}{lrr}

             &                           b   &      a\\ [.4in]

          p-A average (prior)    &      .46   &    .05\\
          A-A average (prior)    &      .46   &    .00\\ [.4in]

          p-A central (prior)    &      .63    &   .00\\
          A-A central (prior)    &      .51    &  -.10\\
\end{tabular}
\end{center}
\end{table}

We also note from the curves for central, i. e., zero impact parameter,
 collisions, Fig 4,
that for a central $AB$ collision the
values of $<n>$ are somewhat different than those of an $AA$ collision
with the same $  (A^{1/3} + B^{1/3}) $ value. 
However, as expected, $A-B$ and $A-A$
average collisions should fall on the $A-A$ curve at the appropriate
\LAB.

 \begin{figure}[h]
 \centerline{\psfig{figure=jpffig4,height=8cm}}
 \caption{}
 \end{figure}

\vss One might guess that a good approximation for the averaging over
all $n$ might be to replace $(N -1)/2$  by $<n>$ in eq. 7. Instead, we will
carry out the averaging to test this hypothesis.

\vss If one examines the distribution of prior collisions, $n$, when the 
average number is $<n>$, the normalized distribution is given 
approximately \cite{tabulation}  by:

\vsp

11)   $F(n) = e^{-n/<n>} / <n> $

\vsp Therefore we can average $S(n)$ over the $n$ distribution to obtain
$S(<n>)$

\vsp 12) $ S(<n>)  = \int e^{\alpha n } e^{-n/<n>}/ <n>  dn $

\vsp This is just

13)  $S(<n>)$ = \begin{large}
$\frac{1}{ (\alpha + 1/<n>)  <n>}  =
 \frac{1}{1 + \alpha <n>} $
\end{large}

Using

14) $1/S(<n>)  = 1 + \alpha <n> $

\begin{figure}[h]
\centerline{\psfig{figure=jpffig5,height=7cm}}
\caption{}
\end{figure}

      To verify this formula,
               we can plot $1/S(<n>) $ vs $\alpha$, which should then
be a straight line whose  slope is the mean number of collisions.
   To do this we  have used  our complete Monte Carlo
calculation for $S(\alpha)$ for average Pb-Pb collisions to determine the 
validity of the approximate calculation of eq. 14.
   We have chosen $\alpha $'s
corresponding with hypothetical absorption
cross-sections from 4.08 to 9.4 mb.

\vsp Fig 5 shows this plot. It is indeed linear with a slope of 0.44.
which is, within error, the same as the value determined from the Monte
Carlo calculation of $<n>$ for a Pb-Pb collision, shown in Fig 3.

\vsp Thus the theoretical assumption for an exponential shape of the
$n$ distribution is apparently a good approximation.

\vsp As a second check, we can examine, for fixed $\alpha$, how the
Monte Carlo calculation of $<n>$ and the approximate calculation of
$<n>$ from eq. 14 depend on $A^{1/3} + B^{1/3} - 2 $ for p-A collisions.
This plot is shown in Fig. 6 showing that both methods agree for a
range of A's from P-C to p-U.

\begin{figure}[h]
\centerline{\psfig{figure=jpffig6,height=10cm}}
\caption{}
\end{figure}

     It is very useful, therefore, to note that one can use the
analytical expression of equation 1) and the parameter $K = .46$ to
calculate effects of absorption whether they are due to disappearance of
the \jp or due to an exponential decrease in yield as the result of
prior energy loss.

\vsp Thus we conclude that an effective $L$ for plotting the suppression
should be 

\vsp
15) $L_{eff} = .46  (A^{1/3} + B^{1/3} - 2) $

\newpage

\centerline{\bf IV. Complete Monte Carlo Calculations of the \jp
Suppression}

\vss
{\bf A. Charmonium Absorption Only   }

\vsp
We first turn to a calculation of the suppression on the assumption that
initial state energy loss can be completely neglected. Fig 7 shows the
effect of varying the absorption cross section obtained from our full
Monte Carlo and based on the same Woods-Saxon distribution used by
M Nardi \cite{privatecomm}.

>From this figure we see that the cross section that fits the data is
greater than 9 mb and thus
 is in disagreement with the cross section obtained from
photoproduction.
\begin{figure}[h]
\centerline{\psfig{figure=jpffig7,height=10cm}}
\caption{}
\end{figure}
 This alone points to the need for energy loss. However
Ref. \cite{privatecomm} has done what is presumably the same calculation
                      and obtain a cross section of 7.6
mb. We believe that this might be due to the following difference in the
details of the calculation:
 In calculating absorption effects it is
important to recognize that the nucleus is composed of finite nucleons
and that there is a counting problem when one attempts calculations
using a smooth density distribution. If one integrates along a path up
to a point within the nucleus and then asks for the probability of a
collision on the way out of the nucleus, the struck nucleon is included in
the absorption path, overestimating the absorption and thus requiring a
smaller absorption cross section. One cannot have the struck nucleon
absorb the charmonium without including this effect in the p-p cross
section as well.
  This effect also occurs in the calculation of the effects of
color screening in high \pt e-p and p-p collisions. There the overcounting
resulted in a claim for observation of color screening. The difference in
the two methods
        is easily demonstrated \cite{physlett92}.

\newpage
{\bf B. Energy Loss Only }
\begin{figure}[h]
\centerline{\psfig{figure=jpffig8a,height=10cm}}
\caption{}
\end{figure}

\vsp In this subsection we show the effects of energy loss with and
without
absorption. Fig 8 shows a plot of average collisions for several
values of \S, corresponding with ones used in present experiments.
The upper lines show the effects of energy loss alone. The lower points
show the effects of changing \S when both absorption and energy loss
are computed.

\newpage
{\bf C. Combining Absorption and Energy Loss }

\begin{figure}[h]
\centerline{\psfig{figure=jpffig9,height=10cm}}
\caption{}
\end{figure}

\vsp In this final section we turn to a complete study of both the
energy loss and absorption effects without the exponential energy loss
approximation. It is obvious from eqs 1) and 2) that energy loss
will produce deviations from pure
exponential behavior of $S$ vs \LAB.
\vsp
 If the energy loss were given by the approximate
relation, namely $e^{-\alpha n}$, we have seen that $S$ falls off almost
exponentially with  $\alpha$. However, as we have seen previously, the
actual energy dependence is more complicated. Unlike the so-called
``scaling factor'' of eq. 1, which depends on \S , the energy loss factor
is more sensitive and depends on $s$, as shown in eq. 2.
 The effect of using the true dependence of the \jp
cross section on energy is to introduce {\it curvature} into the  plot of
$S$ vs the mean number of collisions.
The direction of the curvature is to depress $S$ at high $A$ and low \S
which, as we shall see from our Monte Carlo calculations
 is in agreement with  what the data show.

Fig 9 shows the results using equation 2) with the parameters
\dsdn = .5 + .018 \S and \soc = 6.3 and 7.9 mb. These are reasonable
 choices of the parameters and are consistent with the photoproduction
data on \soc. However, other choices, within limits, give equivalent
results. It is worth pointing out, however, that the data are taken at
different energies and with different experimental arrangements and no
{\it systematic} errors have been provided. We have marked the 
different \S
points so the reader can decide whether slightly larger \soc with a
smaller \dsdn will give a ``better fit'', but we will not do so, in
view of the need for three parameters which are not accurately known.

This simple counting model, incorporating energy loss, accounts
 for the present
data without the need to invoke any ``new physics''.

\newpage
\centerline{\bf VI. Central Collisions}

\begin{figure}[h]
\centerline{\psfig{figure=jpffig10a,height=10cm}}
\caption{}
\end{figure}

\vsp We have seen in section II. that there is an enhancement factor $C$
that multiplies the exponential dependence $S$ on \LAB. This was shown for
the special case of interactions of two lines of nucleons. One can verify
this, of course, using the full exact Monte Carlo. The largest number
of scatters would occur in measurements that trigger on high transverse
energy, so this type of measurement would find deviations from true
exponential behavior. Since high \Et triggers favor central, zero
impact parameter, collisions, we choose to examine central A-A collisions.
Fig 10  shows a plot of central $A-A$ collisions for  \soc = 7.9 mb. The
energy loss is set equal to zero to demonstrate the effect. It is clear
from the full MC calculation that $S$ turns up at large $A$. Once again we
see how simple exponential extrapolations to large $A$ can be misleading.

\newpage
     \centerline{\bf VII. Examination of the \S Approximation  }
\begin{figure}[h]
\centerline{\psfig{figure=jpffig11a,height=10cm}}
\caption{}
\end{figure}

\vsp In section III. we carried out an analytic calculation based on the
approximation in the energy loss equation, going from eq 2 to eq 2'.
To see this effect in more detail, we show in Fig. 11 a comparison of the
two results.  It appears that the approximation is noticeable but produces
only a small effect on the suppression even at high \LAB.

\newpage
\centerline{\bf VIII. Discussion and Future Analyses   }

\vss  Because the soft processes that take place in nuclei are not able
to be handled by perturbative QCD, it is essential to examine how
\jp's are made in a large variety of experiments. It is especially
important to study the energy loss effect since it appears in all particle
production in nuclei.

\vsp
1)
Photon and electron
probes, having no initial state interactions,
should  be used to determine both the \xf dependence of the open
charm cross section and the energy loss of the outgoing \jp. Both are
presently unknown, as is the \jp-nucleon inelastic cross section.
Several incident energies will be needed to unravel these three 
quantities.

2) Feynman x distributions of Drell-Yan dimuons are poorly known. 
 This reaction studies only the initial state
interactions and is the companion to the photo-production experiments.
The Drell-Yan data are the only data that determine \dsdn uniquely, since
there are no final state interactions. This information is needed in
 comparing \jp production relative to Drell-Yan production.
        The Feynman x distributions at \xf = 1 are constrained by a
knowledge of the Glauber coefficients; in fact, R
is given by the probability
that the pair is made in the first collision. Thus the \xf = 1 
measurements are a check on Glauber coefficients and the accuracy of the
nucleon spatial distributions. The \xf A dependence directly checks this
information, which enters into every nuclear reaction.

3) Finally, the Feynman x distributions for the
nuclear \jp cross section relative to p-p and relative
 to the Drell-Yan can be examined and compared for internal consistency.

Even these measurements may not allow one to get a full understanding of
the processes for the following reasons:

1) As pointed out by Hufner et al. \cite{physlett91},
that part of charmonium that is in
the singlet state should be color screened and therefore \soc can easily
be \xf dependent. It cannot be a major effect, since the fall in the
ratio R(\xf) with \xf seems to be dominated by energy loss,
 but any color screening
will contribute to an {\it increase} in the values of $S$.

2) The time evolution of the off-shell charmonium to the on-shell \jp
configuration is poorly known. There are, in fact, no estimates other
than $\Delta E \Delta t$ arguments and no calculations of the time
evolution of the
transition
of a $c-\bar{c} $ color-screened configuration to the \jp on-shell
configuration of charmed quarks and gluons, especially since this is
surely dependent on the density of gluons in normal nuclear matter.
This must be known since the final state interactions of charmonium and
the \jp are not necessarily identical.

      We conclude that careful study of  \jp production in nuclei can
teach us about the interplay between the non-perturbative and perturbative
processes of the \jp, but that claims for ``new physics'' from
total yields have little bearing on these interesting problems that, so 
far, are poorly understood.

\vss \centerline{\bf IX. Conclusions}

\vsp
1) Analytic examination of the \jp suppression, $S$,  shows that $S$ cannot    the
be exactly exponential in \LAB and therefore that such an extrapolation to
large \LAB is invalid.

2) All Feynman \xf p-A data demonstrate that final state absorption of
charmonium cannot be the only contributor to the suppression, so
calculations of suppression in A-B interactions must include the effects
of energy loss in soft scatters prior to charmonium production.

3) Complete calculations using a Woods-Saxon nucleon distribution shows
that there appears to be nothing ``anomalous'' in the suppression
observed in Pb-Pb collisions.

\newpage
     \centerline{ \bf Figures }

\vsp

Fig. 1 \  Measured Feynman x distributions, for the nucleus to proton
target
ratio, $R$,
 of \jp production. (Also shown are fits to the data, including
a crude estimate of the \jp - nucleon  inelastic cross sectiom
in the final state taken from Reference \cite{absorption}.

Fig. 2 \ Schematic of interactions of nucleons in a line: representation
of the collision of a line of $N$ nucleons with a line of $M$ nucleons.

Fig. 3 \  Results of Monte Carlo Woods-Saxon calculation of the mean number
of collisions prior to (or succeeding) an interaction producing charmonium
in p-A and A-A average collisions.

Fig. 4 \ 
        Results of Monte Carlo Woods-Saxon calculation of the mean number
of collisions prior to (or succeeding) an interaction producing charmonium
in $A-A$ central collisions of zero impact parameter as well as for
average $A-A$ collisions. Several $A-B$ collisions show departures from the
straight line slopes, but only for central collisions.

Fig. 5 \ Plot of  1/S = $\alpha + <n>$ from  the exact full Monte-Carlo
calculation, using the values of $<n>$ for Pb-Pb and varying \soc from
4.1 to 9.4 mb. This plot is used to demonstrate the validity of this 
new equation
                 which has been derived using the n distribution of
equation 11.

Fig. 6 \ Comparison of the \LAB  dependence of the
mean number of scatters $<n>_mc$, taken
from the full exact
Monte Carlo calculation of S
with the value, $<n>_{approx}$,
using the derived formula, $ 1/S =   \alpha + <n>_{approx} $,
   for the case \soc = 7.4 mb.

Fig. 7 \ Complete Monte Carlo calculation for pure absorption. The best
fit disagrees with the value of \soc from photoproduction.

Fig. 8 \ Final predicted values of the suppression for different values of
\S for a fixed \soc = 6.3 mb, showing how energy loss has different
suppressions in experiments at different \S.

Fig. 9 \ Comparison of Calculations with Data at Various Values of \S.
The solid line shows the extrapolation of the lower \S data. Calculations
are shown for two values of \soc consistent with the photoproduction
data for a reasonable value of \dsdn (See text.)

Fig. 10 \  Calculated Suppression in Central A-A Collisions for \soc = 7.9
mb and no energy loss, showing that the enhancement factor, C,
shows up in the
full Monte Carlo Calculation.

Fig. 11 \ Effect of using the exact expression for the \S dependence of
the \jp cross section, eq. 2 vs the use of eq 2'. The plot is shown
for \dsdn = .4 + .012s and \soc = 6.9 mb.


\begin{thebibliography}{99}
\bibitem{physlett82} Phys.
Lett. {\bf 119B} 464 (1982); Phys. Rev {\bf D28} 2334 (1983);
These were the experiments that verified, for multiplicity distributions,
the validity of
the Bialas, Bleszynski and
 Czyz  hypothesis (Nucl. Phys. {\bf B111} 461 (1976)
 that the number of participants and not the
number of scatters determined the event structure. It is now appreciated
            that the
basis for the wounded nucleon counting is  the time dilation effect.
\end{thebibliography}
\end{document}